\documentclass[12pt]{article}
\usepackage{epsfig}
\title{Hadronic $\tau$ decay, the renormalization group, analiticity
of the polarization operators and QCD parameters.}
\author{B.V.Geshkenbein \thanks{E-mail address: geshken@heron.itep.ru}\\
Institute of Theoretical and Experimental
Physics,\\
B.Cheremushkinskaya 25, 117218 Moscow,Russia}
\date{}
\begin{document}
\maketitle

\newcommand{\be}{\begin{equation}}
\newcommand{\ee}{\end{equation}}

\def\la{\mathrel{\mathpalette\fun <}}
\def\ga{\mathrel{\mathpalette\fun >}}
\def\fun#1#2{\lower3.6pt\vbox{\baselineskip0pt\lineskip.9pt
\ialign{$\mathsurround=0pt#1\hfil##\hfil$\crcr#2\crcr\sim\crcr}}}

The ALEPH data on hadronic $\tau$-decay is throughly analysed in the
framework of QCD. The perturbative calculations are performed in 1-4-loop
approximation. The analytical properties of the polarization operators are
used in the whole complex $q^2$ plane.
It is shown that the QCD prediction for $R_{\tau}$ agrees with the
measured value $R_{\tau}$ not only for conventional $\Lambda^{conv}_3=(618
\pm 29)~MeV$ but as well as for $\Lambda^{new}_3=(1666 \pm 7)~MeV$.
The polarization operator calculated using the renormgroup has
nonphysical cut $[-\Lambda^2_3, 0]$. If $\Lambda_3 = \Lambda^{conv}_3$, the
contribution of only physical cut is deficient in the explanation of
the ALEPH experiment.  If $\Lambda_3 = \Lambda^{new}_3$ the contribution
of nonphysical cut is very small and only the physical cut explains the
ALEPH experiment. The new sum rules which follow only from analytical
properties of polarization operators are obtained. Basing on the sum rules
obtained, it is shown that there is an essential disagreement between QCD
perturbation theory and the $\tau$-lepton hadronic decay experiment at
conventional value $\Lambda_3$. In the evolution upwards to larger energies
the matching of $r(q^2)$ (Eq.(12)) at the masses $J/\psi$, $\Upsilon$ and $2
m_t$ was performed. The obtained value $\alpha_s(-m^2_z) = 0.141 \pm 0.004$
(at $\Lambda_3 = \Lambda^{new}_3$) differs essentially from conventional
value, but the calculation of the values $R(s) = \frac{\sigma(e^+e^- \to
hadrons)}{\sigma(e^+e^- \to \mu^+\mu^-)}$, ~ $R_l = \frac {\Gamma(Z \to
hadrons)}{\Gamma(Z \to leptons)}$, ~ $\alpha_s(-3~ GeV^2)$, ~ $\alpha_s(-2.5~
GeV^2)$ does not contradict the experiments.


\section {INTRODUCTION}


The purpose of this work is to combine the analyticity requirements of QCD
polarization operators with the renormalization group. This work is the
continuation of works [1-3]. In the work [1]  analytical properties of
polarization operators were used to improve perturbation theory in QCD.
In the works [2,3] the high precision data on hadronic $\tau$-decay obtained
by the ALEPH [4], OPAL [5] and CLEO [6] Collaborations were analyzed in the
framework of QCD.  The analyticity requirements of the QCD polarization
operators follow from the microcausality and the unitarity, therefore we have
no doubts about them.  On the other hand, the calculation according to
renormgroup leads to appearance of nonphysical singularities.  So, the
one-loop calculation gives a nonphysical pole, while in the calculation in a
larger number of loops the pole disappears, but a nonphysical cut appears,
$[-\Lambda^2_3, 0]$. As will be shown, there are only two values of
$\Lambda_3$, such that theoretical predictions of QCD for $R_{\tau, V+A}$
(formulae (23),(24) agree with the experimens [4-6]. These values are the
following: one conventional value $\Lambda^{conv}_3 = (618 \pm 29)~ MeV$ and
the other value of $\Lambda_3$ is $\Lambda^{new}_3 = (1666 \pm 7)~MeV$.
Only in these values of $\Lambda_3$ the predictions of QCD are consistent
with the experiments [4-6]. As far as I know, the value
$\Lambda_3=\Lambda^{new}_3$ was not considered before now \footnote{One can
see this also from Fig.2 of ref.[3], the line obtained in conventional
approach crosses the experimentally allowed strip of $R_{\tau,V+A}$ at two
values of $\Lambda_3$ equal to $\Lambda^{conv}_3$ and $\Lambda^{new}_3$
(in ref.[3] the parameter $\alpha_s(-m^2_{\tau})$ related to $\Lambda_3$
was used). In ref.3 the new value $\Lambda^{new}_3$ was not considered.} If
one simply puts out the nonphysical cut and leaves the conventional value
$\Lambda^{conv}_3$, then the discrepancy between the theory and experiment
will arise. As will be shown, if instead of the conventional value
$\Lambda^{conv}_3$ one chooses the value $\Lambda^{new}_3 = (1565 \pm 193)~
MeV$ then only the physical cut contribution is enough to explain the
experiment of the hadronic $\tau$-decay. \footnote{The errors here and the
following formulae are due only to the error in the measurement of the
value $R_{\tau,V+A}$ (eq.(24))}. It is convenient to introduce the Adler
function (11-13) instead of the polarization operator.  The Adler function
is an analytical function of $q^2$ in the whole  complex $q^2$ plane with a
cut along the positive $q^2$ semiaxes. We will use the renormgroup only for
negative $q^2$, where the value $\alpha_s(q^2)$ is real and positive.  For
other $q^2$ the value $\alpha_s(q^2)$ becomes complex and is obtained by
analytical continuation.

The plan of the paper is the following.

In Section 2 the formulae obtained in paper [3] are transformed to suitable
for this paper form.

In Section 3 the values $\Lambda_3$ are found such that $R_{\tau,V+A} =
3.475 \pm 0.022$ (formula (24)).

In Section 4 we obtain new sum rules for polarization
operator which follow only from analytical properties of the polarization
operator.  These sum rules imply that there is an essential discrepancy
between perturbation theory in QCD and the experiment in hadronic $\tau$
decay at conventional value of $\Lambda_3$.  The power corrections and
instantons cannot eliminate this discrepancy.

Section 5 suggests the method of resolving these discrepancies. At
$\Lambda_3 = (1565 \pm 193)~MeV$ the nonphysical cut gives no contribution
into $R_{\tau, V+A}$ and the physical cut gives the experimentally observed
value $R_{\tau, V+A}$. The previously derived sum rules make no sense if
$\Lambda_3$ is as large.

In Section 6 we go over to larger energies. In the matching procedure we
require continuity of $r(s)$ (Eq.(12)) [1] at masses $J/\psi$,  $\Upsilon$
and $2 m_t$ when going over from $n_f$ to $n_f +1$ flavours. The number of
flavours on the cut is a good quantum number. At every point off the cut all
flavours give a contribution. This follows from the dispersion relation for
the Adler function. The continuity requirement off the cut when changing the
number of flavours violates analytical properties of the polarization
operator. Section 6 presents the results of the calculations in  1-4 loops
for estimation of the precision of the calculations. In
Sections 7-9 we compare the theory with experiment.  In Section 7 the
prediction of the function $R(s)$
is compared with experiments. The
calculated values of the function $R(s)$ are in a very good agreement with
the experiment (Tables 5,6) at $2 \leq \sqrt{s} \leq 47.6 GeV$ except for
the resonance region.

In Section 8 we compare the calculated values of $\alpha_s(-3 GeV^2)$ and
$\alpha_s(- 2.5 GeV^2)$ with the values $\alpha_s(-3 GeV^2)$ and $\alpha_s(-
2.5 GeV^2)$ obtained from the Gross-Llewellyn-Smith sum rule [7] and the
Bjorken sum rule [8]. The results of the calculations are in agreement with
the values obtained from the experiment using these sum rules.

Section 9 presents the calculation of $R_l = \frac{\Gamma(Z \to
hadrons)}{\Gamma(Z \to leptons)}$. The obtained value $R_l$
does not contradict the experiment.

Section 10  is devoted to discussion on the analiticity $\alpha_s(q^2)$. It
is shown that the statement that $\alpha_s(0) = 4 \pi/\beta_0 = 1.396$ is
valid only for one-loop calculations.

\newpage

\section{INITIAL FORMULAE}


In this section the formulae obtained in the paper [3] are transformed to
suitable for this paper form. We will consider three loop approximation
thoroughly. Polarization operators of hadronic currents are defined by the
formula

$$\Pi^J_{\mu\nu}(q)=i~\int e^{iqx}\langle 0 \mid
T~J_{\mu}(x)J^+_{\nu}(0)\mid 0 \rangle d^4 x=
$$
\be = (q_{\mu} q_{\nu} -
g_{\mu \nu} q^2) \Pi^{(1)}_J (q^2) + q_{\mu} q_{\nu} \Pi^{(0)}(q^2)
\ee
where
$$ J = V, A; ~~ V_{\mu} = \bar{u} \gamma_{\mu} d, ~~ A_{\mu} = \bar{u}
\gamma_{\mu} \gamma_5 d
$$
Imaginary parts $\Pi^{(1)}_J, \Pi^{(0)}_J$ are
connected with the measured, so called spectral functions $v_1(s), a_1(s),
a_0(s)$ by the formulae

$$
v_1(s)/a_1(s) = 2 \pi~ Im \Pi^{(1)}_{V/A}(s + i 0)
$$
\be
a_0(s) = 2 \pi~ Im \Pi^{(0)}_A (s + i 0)
\ee
Functions $\Pi^{(1)}_{V/A}$ are analytical functions of $q^2$ with the cuts
$[4m^2_{\pi}, \infty]$ for $\Pi^{(1)}_V$ and $[9m^2_{\pi}, \infty]$ for
$\Pi^{(1)}_A, a_0(s) = 2 \pi^2 f^2_{\pi} \delta (s - m^2_{\pi}), f_{\pi} =
130.7 MeV$.

To get QCD predictions let us use the renormalization group equation in
3-loop approximation [9-10]

\be
q^2 \frac{\partial a}{\partial q^2} = - \beta^{(n_f)}_0 ~a^2 (1 +
b^{(n_f)}_1 a + b^{(n_f)}_2 ~ a^2)
\ee
where
$$
a (q^2) = \frac{\alpha_s(q^2)}{4 \pi}, ~~ \beta^{(n_f)}_0 = 11 - \frac{2}{3}
n_f, ~~ \beta^{(n_f)}_1 = 51 - \frac{19}{3} n_f,
$$
\be
\beta^{(n_f)}_2 = 2857 - \frac{5033}{9} n_f + \frac{325}{27} n^2_f, ~~
b^{(n_f)}_1 = \frac{2 \beta^{(n_f)}_1}{\beta^{(n_f)}_0}, ~~b^{(n_f)}_2 =
\frac{\beta_2^{(n_f)}}{2 \beta^{(n_f)}_0}.
\ee
Here $n_f$ is the number of flavours.

Let us consider for the moment $n_f = 3$ and omit the mark $n_f$. Find
singularities of $a(q^2)$. Integrate equation (3) [3]

\be
\beta_0 ln~\frac{q^2}{\mu^2} = - \int\limits^{a(q^2)}_{a(\mu^2)} \frac{d
a}{a^2(1 + b_1 a + b_2 a^2)}
\ee
Denote the value $q^2$ at which $a(q^2) = \infty$ as $-\Lambda^2_3$.
\footnote{This is the definition of $\Lambda_3$.} Then we get instead of
eq.5:
\be
\beta_0 ln \Biggl (\frac{-q^2}{\Lambda^2_3} \Biggr ) =
\int\limits^{\infty}_{a(q^2)}~ \frac{d a}{a^2(1 + b_1 a + b_2 a^2)} \equiv
f(a)
\ee
According to the known value $a(-m^2_{\tau})$ ~ $\Lambda^2_3$ is
determined by the formula

\be
\Lambda^2_3 = m^2_{\tau} e^{-\frac{f(a(-m^2_{\tau}))}{\beta_0}}
\ee
The integral in the formula (6) is taken and the answer is written as

\be
f(a) = \frac{1}{a} + b_1~ ln a - \frac{1}{\sqrt{b^2_1 - 4b_2}} \Biggl
[\frac{1}{x^2_1}~ln(a - x_1) - \frac{1}{x^2}~ln (a - x_2) \Biggr ]
\ee
where
$$
x_{1, 2} = \frac{- b_1 \pm \sqrt{b^2_1 - 4 b_2}}{2 b_2} \eqno{(8a)}
$$
At $\alpha_s (-m^2_{\tau}) = 0.355$~ [3]  we obtain $\Lambda^2_3 = 0.394
GeV^2$.

The expansion of the function $f(a)$ in the Taylor series  at large $a$ over
$1/a$ is of the form

\be
f(a) = \frac{1}{3b_2}~ \frac{1}{a^3} - \frac{b_1}{4 b^2_2}~ \frac{1}{a^4} +
\frac{(b^2_1 - b_2)}{5 b^3_2}~ \frac{1}{a^5} + 0(\frac{1}{a^6})
\ee
It follows from eqs.(6-9) that the singularity $a(q^2)$ at $q^2 \to
- \Lambda^2_3$ has the form [3]

\be
a(q^2) = \Biggl ( -\frac{2 \Lambda^2_3}{3 \beta_2 (q^2 + \Lambda^2_3)}
\Biggr )^{1/3}
\ee
Since for massless quarks the contributions of $V$ and $A$ coincide, we
will omit in all formulae the mark $J$. Introduce the Adler function

\be
D(q^2) = -q^2 \frac{d \Pi (q^2)}{d q^2}
\ee
It is convenient to write for
three flavours

$$
 D(q^2) = 3 (1 + d(q^2)), ~~ R(q^2) = 3(1 + r(q^2)),
~ \Pi(q^2) = 3 (- ln (- q^2/\mu^2) + p(q^2))
$$
\be
r(q^2) =
\frac{1}{\pi} Im p (q^2) = \frac{1}{2 \pi i} [p (q^2 + i 0) - p(q^2 - i0) ]
\ee
In 3-loop approximation for $\overline{MS}$ renormalization scheme
function $d(q^2)$ for negative $q^2$ is written as [11]:
\be
d(q^2) = 4
a(q^2)(1 + 4d^{(n_f)} _1 a(q^2) + 16 d^{(n_f)}_2 a(q^2)^2)
\ee
where
$$
d^{(n_f)}_1 =
1.9857 - 0.1153 n_f
$$
\be
d^{(n_f)}_2 = 18.244 - 4.216 n_f + 0.086 n^2_f
\ee
Hereafter we will follow [3]

\be
d(q^2) = -q^2 \frac{dp(q^2)}{d q^2}
\ee

\be
p(q^2) = - \int~ \frac{d (q^2)}{q^2} dq^2
\ee
Using formula (3)

\be
\frac{d q^2}{q^2} = - \frac{d a}{\beta_0 a^2(1 + b_1 a + b_2 a^2)}
\ee
we get
 for the function $p(q^2)$ the expression

\be
p(q^2) = \frac{1}{\beta_0}~ \int~ \frac{d(a) da}{a^2(1 + b_1 a + b_2 a^2)} =
\frac{1}{\beta_0 b_2}~ \int~ \frac{d(a) da}{a^2 (a - x_1) (a - x_2)}
\ee
After taking the integral (18) we get

\be
p(a) = \frac{4}{\beta_0} ln a +\frac{4}{\beta_0 \sqrt{b^2_1 - 4 b_2}}
\left \{ (\frac{1}{x_1} + 4 d_1 + 16 d_2 x_1) ln (a - x_1) - (\frac{1}{x_2}
+ 4 d_1 + 16 d_2 x_2) ln (a - x_2) \right \}
\ee
$x_1$, $x_2$ are determined
by formula (8a).

The polarization operator is an analytical function with the cut $[0,
\infty]$. The polarization operator calculated in 3-loop approximation has
the physical cut $0 < q^2 < \infty$ and nonphysical one $-
\Lambda^2_3 < q^2 < 0$.  The contribution of the physical cut
in the value $R^{S=0}_{\tau, V+A}$ is equal to

\be
R^{QCD}_{\tau,V+A} \Biggl \vert_{phys.cut} = 6 \vert V_{ud} \vert^2
S_{EW} \int\limits^{m^2_{\tau}}_{0} \frac{ds}{m^2_{\tau}} (1 -
\frac{s}{m^2_{\tau}})^2 (1 + 2 \frac{s}{m^2_{\tau}})(1 + r(s)) ds + \Delta
R^{(0)}_{\tau}
\ee
where $\vert V_{ud} \vert = 0.9735 \pm 0.0008$ is the
Cabibbo-Kobayashi-Maskawa matrix element [12], $S_{EW} = 1.0194 \pm
0.040$ is the contribution of electroweak corrections [13].

\be
\Delta R^{(0)}_{\tau} = - 24 \pi^2 \frac{f^2_{\pi}m^2_{\pi}}{m^2_{\tau}}
 = -0.008.  \ee is a small correction from the pion pole [3]. The
nonphysical cut contribution is equal to

\be
R^{(QCD)}_{\tau,V+A} \Biggl \vert_{nonphysical cut} =6 \vert V_{ud} \vert^2
S_{EW} \int\limits^{0}_{-\Lambda^2_3}~\frac{ds}{m^2_{\tau}} (1 -
\frac{s}{m^2_{\tau}})^2 (1 + 2 \frac{s}{m^2_{\tau}}) (1+r(s)) ds
\ee


\section{Finding of the value $\Lambda_3$}


The value measured in the experiment is

$$
R_{\tau, V+A} = \frac{B(\tau \to \nu_{\tau} + hadrons,~ S=0)}{B(\tau \to
e^- \bar{\nu}_e \nu_{\tau})} =
$$
\be
= 6 \vert V_{ud} \vert^2 S_{EW}
\int\limits^{m^2_{\tau}}_{0} \frac{ds}{m^2_{\tau}} (1 -
\frac{s}{m^2_{\tau}})^2  \Biggl [(1 + 2 \frac{s}{m^2_{\tau}})(v_1(s) +
a_1(s)+ a_0(s)) - 2 \frac{s}{m^2_{\tau}} a_0 (s) \Biggr ]
\ee
For the value
$R_{\tau, V+A}$ , the ALEPH [4], OPAL [5] and CLEO [6] Collaborations had
obtained

\be
R_{\tau, V+A} = 3.475 \pm 0.022
\ee
Here new value of $R_{\tau, S}$ [14,15] is taken into account.

\be
R_{\tau, S} = 0.161 \pm 0.007
\ee

The convenient way to calculate the $R_{\tau}$ in QCD is to transform the
 integral in the complex $s$ plane [16-19] around the circle $\vert s \vert
= m^2_{\tau}$ and thus getting a satisfactory agreement with the experiment.

\be
R_{\tau,V+A} = 6 \pi i\vert V_{ud} \vert^2 S_{EW} \oint_{\vert s \vert =
m^2_{\tau}}~ \frac{ds}{m^2_{\tau}} (1 - \frac{s}{m^2_{\tau}})^2 (1 + 2
\frac{s}{m^2_{\tau}})(1 + 2 \frac{s}{m^2_{\tau}})  \Pi(s) + \Delta
R^{(0)}_{\tau}
\ee
There are two values of $\Lambda_3$, such that $R_{\tau,V+A} = 3.475 \pm
0.022$ (Eq.24).  These values are: conventional value

\be
\Lambda^{conv}_3 = 618 \pm 29 MeV
\ee
and the alternative new value
\be
\Lambda^{new}_3 = 1666 \pm 7 MeV
\ee

We can calculate $R^{QCD}_{\tau,V+A}$  also with the help of formulae
(20,22). At $\Lambda_3 = 618 \pm 29 MeV$
\be
R^{QCD}_{\tau,V+A} \Biggl \vert _{phys.cut} = 3.305 \pm 0.008
\ee
and
\be
R^{QCD}_{\tau,V+A} \Biggl \vert_{nonphys.cut} = 0.162 \pm 0.015
\ee
The sum of integrals on the physical and nonphysical cuts is equal to the
integral over the circle  (it follows from Cauchy theorem) and
coincides with the measured value $R_{\tau,V+A}$ (24). The contribution of
only one physical cut is insufficient to explain the experiment.

If $\Lambda_3 = 1666 \pm 7 MeV$

\be
R^{QCD}_{\tau,V+A} \Biggl \vert_{phys.cut} = 3.480 \pm 0.0007
\ee
and
\be
R^{QCD}_{\tau,V+A} \Biggl \vert_{phys.cut} = 0.0129 \pm 0.0024
\ee
If $\Lambda_3 = \Lambda^{conv.}_3$, the nonphysical cut must be taken into
account to avoid a discrepancy with the experiment. If $\Lambda_3 =
\Lambda^{new}_3$, there is two possibilities. It is possible to omit the
contribution of the nonphysical cut and to satisfy the requirements of
microcausality and unitarity. Alternatively, the contribution of the
nonphysical cut is taken into account and the requirement of microcausality
and unitarity will be satisfied only in a future comprehensive theory.

In refs.[4] $R_{\tau,V}$ and $R_{\tau,A}$ are measured separately.

\be
R_{\tau,V} = 1.775 \pm 0.017
\ee

\be
R_{\tau,A}= 1.717 \pm 0.018
\ee
The values $R_{\tau,V}$ and $R_{\tau,A}$ have been corrected taking into
account papers [14,15].

In QCD one should have for massless $u$ and $d$
quarks

\be
R^{(QCD)}_{\tau,V} = R^{(QCD)}_{\tau,A}
\ee
The results of the experiments (33, 34) contradict the formula (35).
This contradiction was resolved in the paper [2].

\section{NEW SUM RULES FOR POLARIZATION OPERATORS.}


 To derive the sum rule, let us consider the integral over closed contour
from the function $W(s) \Pi(s)$, where $\Pi(s)$ is one of the functions
$\Pi^{(1)}_V(s)$,~ $\Pi^{(1)}_A(s)$,~ $\Pi^{(1)}_V(s) + \Pi^{(1)}_A(s)$,~
$\Pi^{(1)}_V(s) - \Pi^{(1)}_A(s)$, and $W(s)$ is the weight analytical
function, which will be chosen later. As a contour, we choose  that which
contains the upper and lower edges of the cut from $s_1$ to $s_2$ and of two
circles with radii $s_1$ and $s_2$ (see Fig.1). Let us choose the values
$s_1 = 0.6, 0.8, ... 2 GeV^2$ and the values $s_2 = s_1 + 0.2,~ s_1 + 0.4,
... 3 GeV^2$.  The integral considered through Cauchy theorem is zero. It is
does not contain the contribution of power corrections \footnote{We ignored
the $\alpha_s$ corrections to the condensates. The condensates without
$\alpha_s$ corrections are the poles off the contour of integration} and
nonphysical cut.  As a weight function we choose


\begin{figure}[tb] \hspace{30mm}
\epsfig{file=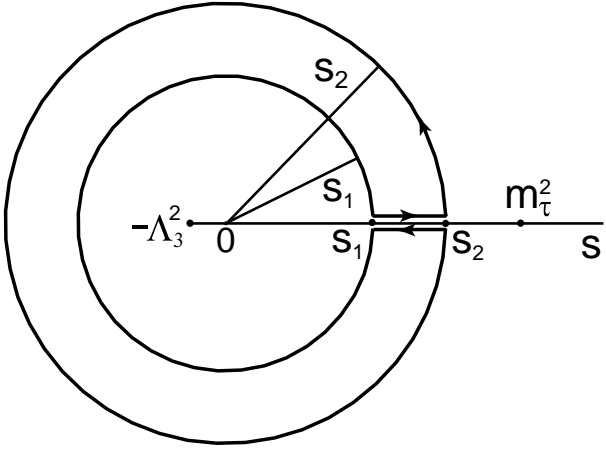, width=80mm}
\caption{Contour integral}
\end{figure}

\be
W(s) = (s - s_1)(s_2 - s)
\ee
The sum of integrals over cut edges is $2i~\int\limits^{s_2}_{s_1}~ W(s) Im
\Pi(s) ds$. This sum is equal to the sum of integrals with inverse sign over
the circles, for which owing to  that the weight function vanishes at the
points $s_1$ and $s_2$, one may take $\Pi^{(QCD)}(s)$ instead of the
true value $\Pi(s)$.

Making use of the analytical properties of $\Pi^{(QCD)}(s)$, let us transform
the sum of integrals over circles into the integral from $Im \Pi^{(QCD)}(s)$
over the cut from $s_1$  to $s_2$. Finally, we obtain the following sum rule

\be
\int\limits^{s_2}_{s_1}~W(s) Im \Pi(s) ds = \int\limits^{s_2}_{s_1}~W(s) Im
\Pi^{(QCD)}(s) ds
\ee
To compare QCD predictions with experiment, let us introduce the notations

\be
U_B = \int\limits^{s_2}_{s_1}~ W(s) Im \Pi_B (s) ds/
\int\limits^{s_2}_{s_1}~W(s) Im \Pi^{(QCD)}_B (s) ds
\ee
where $B = V, A,~ V + A$. The results of the calculations of $U_B$ are
given in Table 1.

\newpage

\begin{center}
Table 1.

\vspace{3mm}
Comparison of the sum rules (37) with the ALEPH experiment.

$U_B$ is given by Eq.(38). $Im \Pi_B(s)$ is obtained from the ALEPH
experimental data. $Im \Pi^{(QCD)}_B(s)$ is calculated by three-loop
approximation of QCD at $\Lambda_3 = 618 MeV$. $s_1$, $s_2$ are given in
$GeV^2$.
\end{center}

\vspace{5mm}

\noindent
\begin{tabular}{|l|l|l|l|l|l|l|l|l|l|l|l|}
\hline $s_2$ & 1.2 & 1.4 & 1.6 & 1.8 & 2 & 2.2 & 2.4 & 2.6 &
2.8 & 3 &\multicolumn{1}{|l|}{~~~}\\ \cline{1-11} $U_V$ & 0.552 &
0.489 & 0.479 & 0.498 & 0.534 & 0.592 & 0.668 & 0.745 & 0.816 &
0.881 & \multicolumn{1}{|l|}{~~~}
\\ \cline{1-11} $U_A$ & 1.169 & 1.398 & 1.502 & 1.514 & 1.465 &
1.389 & 1.309 & 1.229 & 1.161 & 1.107&
\multicolumn{1}{|l|}{$s_1=0.8$}
\\ \cline{1-11} $U_{V+A}$ & 0.861 & 0.942 & 0.988 & 1.003 & 0.995
& 0.985 & 0.982 & 0.981 & 0.982 & 0.984
&\multicolumn{1}{|l|}{~~~}\\ \hline $s_2$ & 1.2 & 1.4 & 1.6
& 1.8 & 2 & 2.2 & 2.4 & 2.6 & 2.8 & 3& \\ \cline{1-11} $u_V$ &
0.438 & 0.425 & 0.451 & 0.494 & 0.546 & 0.62 & 0.71 & 0.796 &
0.871 & 0.939&\\ \cline{1-11} $u_A$ & 1.491 & 1.629 & 1.649 &
1.594 & 1.493 & 1.383 & 1.279 & 1.889 & 1.114 & 1.059 &
$s_1=1$\\ \cline{1-11} $u_{V+A}$ & 0.967 & 1.024 &
1.047 & 1.039 & 1.015 & 0.995 & 0.998 & 0.986 & 0.986 & 0.988 &\\
\hline  $s_2$ & 1.4 & 1.6 & 1.8 & 2 & 2.2 & 2.4 & 2.6 & 2.8
& 3 &\multicolumn{2}{|c|}{~~~}\\ \cline{1-10} $u_V$ & 0.429 &
0.477 & 0.531 & 0.592 & 0.677 & 0.778 & 0.869 & 0. 945 & 1.011 &
\multicolumn{2}{|c|}{~~~}\\ \cline{1-10} $u_A$ & 1.712 & 1.668 &
1.563 & 1.427 & 1.299 & 1.188 & 1.098 & 1.027 & 0.978 &
\multicolumn{2}{|c|}{$s_1=1.2$}
\\ \cline{1-10} $u_{V+A}$ & 1.065 & 1.069 & 1.042 & 1.003 & 0.98 & 0.976
& 0.976 & 0.979 & 0.983 & \multicolumn{2}{|c|}{~~~}\\ \hline
$s_2$ & 1.6 & 1.8 & 2 & 2.2 & 2.4 & 2.6 & 2.8 & 3 &
\multicolumn{3}{|c|}{~~~}\\ \cline{1-9} $u_V$ & 0.533 & 0.586 &
0.65 & 0.749 & 0.862 & 0.956 & 1.028 & 1.091 &
\multicolumn{3}{|c|}{~~~}\\ \cline{1-9} $u_A$ & 1.601 & 1.456 &
1.302 & 1.172 & 1.067 & 0.986 & 0.927 & 0.891 &
\multicolumn{3}{|c|}{$s_1=1.4$}\\ \cline{1-9}
$u_{V+A}$ & 1.063 & 1.015 & 0.968 & 0.951 & 0.956 & 0.963 & 0.970
& 0.978&\multicolumn{3}{|c|}{~~~}
\\ \hline $s_2$ & 1.8 & 2 & 2.2 & 2.4 & 2.6 & 2.8 & 3&
\multicolumn{4}{|c|}{~~~}\\ \cline{1-8} $u_V$ & 0.642 & 0.715 &
0.837 & 0.962 & 1.053 & 1.118 & 1.173 &
\multicolumn{4}{|c|}{~~~}\\ \cline{1-8} $u_A$ & 1.299 & 1.146 &
1.032 & 0.944 & 0.88 & 0.833 & 0.816&
\multicolumn{4}{|c|}{$s_1=1.6$}\\ \cline{1-8}
$u_{V+A}$ & 0.964 & 0.922 & 0.924 & 0.944 & 0.959 & 0.97 &
0.98&\multicolumn{4}{|c|}{~~~}\\ \hline $s_2$ & 2 & 2.2 &
2.4. & 2.6 & 2.8 & 3 & \multicolumn{5}{|c|}{~~~}\\ \cline{1-7}
$u_V$ & 0.788 & 0.956 & 1.086 & 1.161 & 1.209 & 1.252 &
\multicolumn{5}{|c|}{~~~}\\ \cline{1-7} $u_A$ & 1.006 & 0.918 &
0.847 & 0.799 & 0.77 & 0.765 &
\multicolumn{5}{|c|}{$s_1=1.8$}\\ \cline{1-7}
$u_{V+A}$ & 0.884 & 0.923 & 0.958 & 0.974 & 0.983 & 0.922
  & \multicolumn{5}{|c|}{~~~}\\ \hline
 $s_2$ & 2.2 & 2.4 & 2.6 & 2.8 & 3 & \multicolumn{6}{|c|}{~~~}\\
 \cline{1-6} $u_V$ & 1.133 &
1.217 & 1.255 & 1.281 & 1.312& \multicolumn{6}{|c|}{~~~}\\
\cline{1-6} $u_A$ & 0.833 & 0.778 & 0.744 & 0.728 &
0.738&\multicolumn{6}{|c|}{$s_1=2$} \\ \cline{1-6}
$u_{V+A}$ & 0.972 & 0.993 & 0.996 & 0.998 &
1.005&\multicolumn{6}{|c|}{~~~}\\ \hline
\end{tabular}

\vspace{5mm}

It is seen from Table 1 that QCD predictions agree with experiment for $V +
A$ and disagree with experiments  for $V$ and $A$ separately. The results do
not change if one takes the weight function of the form $(s - s_1)^n(s_2 -
s)^n$, $n = 2,3,...10$.

Let us consider $V - A$. In this case $Im
(\Pi^{(QCD)}_V(s) - \Pi^{(QCD)}_A(s)) = 0$, ~ while $Im(\Pi_V(s) - \Pi_A (s))
\not= 0$. Try to eliminate this disagreement with the help of
instantons.

The instanton contribution into $\Pi^{(1)}_V(s) - \Pi^{(1)}_A(s)$ in the
model considered in [3] is given by the formula (39) [3]:

\be
\Pi^{(1)}_{V, inst.} (s) -\Pi^{(1)}_{A, inst}(s) = \int\limits^{\infty}_{0} d
\rho n (\rho) \Biggl [ -\frac{4}{s^2} - \frac{4 \rho^2}{s}~ K^2_1 (\rho
\sqrt{-s}) \Biggr ]
\ee
$K_1$ is the Macdonald function. Introduce the notations

\be
L_{Exp} = \int\limits^{s_2}_{s_1}~ W(s) Im (\Pi^{(1)}_V(s) - \Pi^{(1)}_A(s))
ds
\ee

\be
L_{inst} = \int\limits^{s_2}_{s_1}~ W(s) Im (\Pi^{(1)}_{V, inst} (s) -
\Pi^{(1)}_{A, inst}(s)) ds
\ee

The results of $L_{Exp}$ and $L_{inst}$ calculations are given in Table 2
for $n(\rho)=n_0\delta(\rho-\rho_0)$ $\rho_0 = 1.7 GeV^{-1}$,~ $n_0 = 1.5
\cdot 10^{-3} GeV^4$.

\vspace{5mm}

\begin{center}
{\bf Table 2.}  $s_1 = 0.8 GeV^2$.
\end{center}

\bigskip

\hspace{2.1cm}
\begin{tabular}{|c|c|c|c|c|c|c|}
\hline
$s_2/GeV^2$ & 1 & 1.2 & 1.4 & 1.6 & 1.8\\
\hline
$L_{Exp}$ &-0.00042& -0.022 & -0.107 & -0.286 & -0.554\\
\hline
$L_{inst}$ & 0.000144 &
0.00089 & 0.00234 & 0.00427 & 0.00634\\
\hline
$L_{Exp}/L_{inst}$ & -2.907 & -24.147 & -45.8 & -66.82 & -87.34\\
\hline
\end{tabular}

\vspace{3mm}
\hspace{2cm}
\begin{tabular}{|c|c|c|c|c|c|c|}
\hline
$s_2/GeV^2$ & 2 & 2.2 & 2.4 & 2.6 & 2.8 & 3\\
\hline
$L_{Exp}$ & -0.878 & -1.198 & -1.442 & -1.572 & -1.563 & -1.402\\
\hline
$L_{inst}$ & 0.00815 & 0.00933 & 0.00955 & 0.00853 & 0.00619 & 0.00209\\
\hline
$L_{Exp}/L_{inst}$ & -107.7 & -128.7 & -151.1 & -184.2 & -256.7 & -670.7\\
\hline
\end{tabular}


\vspace{3mm}

It is seen from Table 2 that instantons (in the model under consideration)
cannot eliminate the disagreement between QCD theory and experiment.

\section{NEW QCD PARAMETERS AND ELIMINATION\\
OF CONTRADICTIONS}

\vspace{5mm}
In my opinion, the only possible way to resolve the discrepancy which
follows from the sum rules (37)  is to change the conventional value
$\Lambda^{conv}_3 \sim 600 MeV$ by $\Lambda^{new}_3 \sim 1600 MeV$.
Because $s_1$ must be larger than $\Lambda^2_3 \sim 2.5 GeV^2$, the sum
rules (38) become meaningless.  At $\Lambda^{new}_3 \sim 1600 MeV$ we
have a possibility to fulfill the requirement of microcausality and
unitarity to omit the nonphysical cut.

The contribution of the physical cut

\be
R^{QCD, S=0}_{\tau, V+A} = 6 \vert V_{ud} \vert^2 S_{EW}
\int\limits^{m^2_{\tau}}_{0}~ \frac{ds}{m^2_{\tau}} (1 -
\frac{s}{m^2_{\tau}})^2 (1 + \frac{2s}{m^2_{\tau}}) (1 + r(s)) ds = 3.483
\pm 0.022
\ee
agrees with experiment (24) at
\be
\Lambda_3 = \Lambda^{new}_3 = (1565 \pm 193) MeV
\ee
The value $\Lambda^{new}_3$ in eq.(43) differs from $\Lambda^{new}_3$ in
eq.(28) since in (28) we take into account the contribution of the
nonphysical cut. It is my belief, that the nonphysical cut must be absent.
In spite of that the nonphysical cut  is a consequence of 3-loop in p QCD,
if we are able to eliminate this drawback, we must do it. But at $\Lambda_3
\sim 600 MeV$ we cannot do it while at $\Lambda_3 \sim 1600 MeV$ we can. In
what follows we put $\Lambda_3 = (1565 \pm 193) MeV$ and omit the
nonphysical cut contribuion.


\section{TRANSITIONS TO A LARGER NUMBER OF FLAVOURS}


Let us introduce the notations

\be
f_{n_f} (a) = \frac{1}{a} + b^{(n_f)}_1~ ln a -
\frac{1}{\sqrt{b^{(n_f)^2}_1 - 4 b^{(n_f)}_2}}\Biggl
[\frac{1}{x^{(n_f)^2}_1}~ln (a - x^{(n_f)}_1)~- \frac{1}{x^{(n_f)^2}_2} ln
(a - x^{(n_f)}_2) \Biggr ], ~~ n_f \leq 5
\ee

\be
f_{n_f}(a) = \frac{1}{a} + b^{n_f)}_1 ~ ln a - \frac{1}{\sqrt{b^{(n_f)^2} -
4b^{(n_f)}_2}} \Biggl [ \frac{1}{x^{(n_f)^2}_1} ~ln (a - x^{(n_f)}_1) -
\frac{1}{x^{(n_f)^2}} ~ln (x^{(n_f)}_2 - a) \Biggr ], ~~ n_f = 6 \footnote{The sign of the argument
in the third logarithm is changed for that to remain on the physical sheet.}
\ee

where
\be
x^{(n_f)}_{1,2} = \frac{-b^{(n_f)}_1 \pm \sqrt{b^{(n_f)^2}_1 - 4
b^{(n_f)}_2}}{2 b^{(n_f)}_2}
\ee

$$
x^{(3)}_1 = - 0.0497 + 0.107 i, ~~ x^{(3)}_2 = x^{(3)*}_1, ~~ x^{(4)}_1 =
-0.0632+0.1296 i, ~~ x^{(4)}_2 = x^{(4)*}_1,
$$
$$
x^{(5)}_1 = -0.107 + 0.176 i, ~~ x^{(5)}_2 = x^{(5)*}_1, ~~ x^{(6)}_1 =
-0.213, ~~ x^{(6)}_2 = 1.013
$$
The value $a(q^2) $ is found by numerical solution of the
equation

\be
\beta^{(n_f)}_0~ln \Biggl ( - \frac{q^2 + i 0}{\Lambda^2_f} \Biggr ) =
f_{n_f}(a),~~ q^2 > 0, ~~ n_f = 3,4,5,6
\ee
The function $r(s)$ can be obtained with the help of Eqs.(19,12).

In the evolution upwards to larger energies the matching of $r(q^2)$ at the
masses $J/\psi$, $\Upsilon$ and $2m_t$ is performed.

There are three alternatives:

1) The nonphysical cut is absent $\Lambda^{new}_3 = (1565 \pm 193) MeV$.
The Adler function $d(q^2)$ may be written in
the form
\be
d(q^2) = d^{(3)} (q^2) + d^{(4)} (q^2) + d^{(5)} (q^2) + d^{(6)} (q^2)
\ee
where

\be
d^{(3)} (q^2) = - q^2 \int\limits^{m^2_{\psi}}_{0} \frac{r_3 (q^{\prime 2})
dq^{\prime 2}}{(q^{\prime 2} - q^2)^2}
\ee
is the contribution of the part of
the cut with 3 flavours into Adler function. Similarly,

\be
d^{(4)} (q^2) = - q^2~ \int\limits^{m^2_{\Upsilon}}_{m^2_{\psi}}~
\frac{r_4(q^{\prime 2}) dq^{\prime 2}}{(q^{\prime 2} - q^2)^2}
\ee
is the contribution of the part of the cut with 4 flavours into Adler
function

\be
d^{(5)}(q^2) = - q^2~ \int\limits^{4 m^2_t}_{m^2_{\Upsilon}}~ \frac{r_5
(q^{\prime 2}) dq^{\prime 2}}{(q^{\prime 2} - q^2)^2}
\ee
is the contribution of the part of the cut with 5 flavours into Adler
 function
\be
d^{(6)}(q^2) = - q^2~ \int\limits^{\infty}_{4m^2_t}~
\frac{r_6 (q^{\prime 2}) dq^{\prime 2}}{(q^{\prime 2} - q^2)^2}
\ee
is the contribution of
the part of the cut with 6 flavours into Adler function.

The number of flavours for $r(q^2)$ on the cut is a certain number in
contrast to the number of flavours  at the point of the complex plane $q^2$
off the cut. Let us consider $q^2 = - m^2_Z$ and find $\alpha_s(-m^2_Z)$.

Return to formula (13).The coefficients $d_1$ and $d_2$ in (13) are defined
for a certain number of flavours.

Introduce
\be
d^{(av)}_1 (-m^2_Z) = (d^{(3)}_1 d^{(3)} (-m^2_Z) + d^{(4)}_1 d^{(4)}
(-m^2_Z) + d^{(5)}_1 d^{(5)} (-m^2_Z) + d^{(6)}_1 d^{(6)}
(-m^2_Z))/d(-m^2_Z)
\ee

\be
d^{(av)}_2 (-m^2_Z) = (d^{(3)}_2 d^{(3)} (-m^2_Z) + d^{(4)}_2 d^{(4)}
(-m^2_Z) + d^{(5)}_2 d^{(5)} (-m^2_Z) + d^{(6)}_2 d^{(6)}
(-m^2_Z))/d(-m^2_Z)
\ee

The values $d^{(n_f)} (-m^2_Z)$ have been calculated.

$$
d^{(3)} (-m^2_Z) = 0.000169, ~~ d^{(4)} (-m^2_Z) = 0.000823, ~~ d^{(5)}
(-m^2_Z) = 0.0432
$$
\be
d^{(6)} (-m^2_Z) = 0.00218, ~~ d(-m^2_Z) = 0.0464
\ee
Formula (13) is replaced by

\be
d(-m^2_Z) = 4 a(-m^2_Z)(1 + 4 d^{(av)}_1 (-m^2_Z) a(-m^2_Z) + 16 d^{(av)}_2
(-m^2_Z) a^2 (-m^2_Z))
\ee
The equation (49) can be solved for $a(-m^2_Z)$

\be
\alpha_s(-m^2_Z) = 4 \pi~a (-m^2_Z) = 0.142 \pm 0.004
\ee
The value $\alpha_s(m^2_Z + i 0)$ is evaluated from (6,8).
$$
\alpha_s(m^2_Z + i 0) = 0.131 \pm 0.003 + (0.037 \pm 0.002) i
$$
\be
\vert \alpha_s (m^2_Z + i 0) \vert = 0.136 \pm 0.004
\ee

In a similar way $\alpha_s(q^2)$ can be calculated at arbitrary $q^2$. The
values of $\alpha_s$ at the interesting points are given in Table 3.

\newpage

\begin{center}
Table 3.
\end{center}

\vspace{5mm}

The calculation of $\alpha_s(q^2)$ at different  $q^2$ in approximation of
1-4 loops. The matching of $r(q^2)$ at the masses of $J/\psi$, $\Upsilon$ and
$2m_t$ is performed. The contribution of nonphysical cut is omitted.

\vspace{5mm}

\begin{tabular}{|l|l|l|l|l|l|l|}
\hline
Approximation & $\Lambda_3/MeV$ & $\Lambda_4/MeV$ & $\Lambda_5/MeV$ &
$\Lambda_6/MeV$ & $\alpha_s(0)$ & $\alpha_s(-m^2_{\tau})$\\
\hline
One loop & $618\pm59$ & $508\pm51$ & $377\pm 40$ & $191\pm 22$ & 1.396 &
$0.469\pm 0.018$\\
\hline
Two loops & $1192\pm 136$ & $956\pm 113$ & $678\pm 86$ & $301\pm 42$ &
$0.895\pm 0.001$ & $0.387\pm 0.015$\\
\hline
Three loops & $1565\pm 193$ &
$1257\pm 158$ & $872\pm 119$ & $312\pm 47$ & $0.749\pm 0.001$ & $0.379\pm
0.013$\\
\hline
Four loops & $1862\pm 230$ & $1503\pm 189$ & $1064\pm 145$ &
$202\pm 31$ & $0.749\pm 0.001$ & $0.379\pm 0.013$\\
\hline
\end{tabular}

\vspace{5mm}
\begin{tabular}{|l|l|l|l|}
\hline
Approximation & $\alpha_s(m^2_{\tau}+i 0)$ & $\alpha_s(-m^2_{\psi})$ &
$\alpha_s(m^2_{\psi}-0+i 0)$\\
\hline
One loop & $0.356\pm0.019+(0.306\pm0.017) i$ & $0.377\pm0.014$ &
$0.276\pm 0.010+(0.216\pm 0.013)i$\\
\hline
Two loops & $0.355\pm 0.014+(0.270\pm0.014) i$ & $0.332\pm 0.013$ & $0.265\pm
0.003+(0.200\pm 0.014) i$\\
\hline
Three loops & $0.365\pm 0.023+(0.264\pm 0.013) i$ & $0.322\pm 0.009$ &
$0.272\pm 0.015\pm (0.201\pm 0.014) i$\\
\hline
Four loops & $0.363\pm 0.013+(0.263\pm 0.013) i$ & $0.326\pm 0.012$ &
$0.273\pm 0.016\pm (0.202\pm 0.014) i$ \\
\hline
\end{tabular}

\vspace{5mm}

\begin{tabular}{|l|l|l|l|}
\hline
Approximation & $\alpha(m^2_{\psi}+0+i0)$ & $\alpha_s(-m^2_{\Upsilon})$ &
$\alpha_s(m^2_{\Upsilon}-0+i0)$\\
\hline
One loop & $0.291\pm 0.012+(0.206\pm 0.013) i$ & $0.254\pm 0.008$ & $0.206\pm
0.005+(0.107\pm 006) i$ \\
\hline
Two loops & $0.284\pm 0.002+(0.193\pm 0.014) i$ & $0.234\pm 0.009$ &
$0.192\pm 0.004+(0.100\pm 007) i$\\
\hline
Three loops & $0.293\pm
0.017+(0.199\pm 0.015) i$ & $0.237\pm 0.009$ & $0.194\pm 0.006+(0.106\pm
0.008) i$\\
\hline
Four loops & $0.294\pm 0.018+(0.199\pm 0.015) i$ & $0.237\pm
0.009$ & $0.194\pm 0.007+(0.104\pm 0.007) i$\\
\hline
\end{tabular}

\vspace{5mm}
\begin{tabular}{|l|l|l|l|}
\hline
Approximation & $\alpha_s(m^2_{\Upsilon}+0 + i 0)$ & $\alpha_s(-m^2_z)$ &
$\alpha_s(\psi m^2_z+i 0)$\\
\hline
One loop & $0.211\pm 0.006+(0.100\pm 0.005) i$ & $0.142\pm 0.003$ & $0.141\pm
0.006+(0.039\pm 0.002) i$\\
\hline
Two loops & $0.199\pm 0.005+(0.093\pm 0.007) i$ & $0.137\pm 0.002$ &
$0.129\pm 0.003+(0.036\pm 0.002) i$\\
\hline
Three loops & $0.203\pm 0.007+(0.098\pm 0.007) i$ & $0.142\pm 0.004$ &
$0.131\pm 0.003+(0.037\pm 0.002) i$\\
\hline
Four loops & $0.204\pm 0.008+(0.099\pm 0.007) i$ & $0.141\pm 0.004$ &
$0.131\pm 0.003+(0.038\pm 0.002) i$ \\
\hline
\end{tabular}

\vspace{5mm}

\begin{tabular}{|l|l|l|}
\hline
Approximation & $r(m^2_z)$ & $R_l$ \\
\hline
One loop & $0.0462\pm 0.0009$ & $20.856\pm 0.017$\\
\hline
Two loops & $0.0458\pm 0.0011$ & $20.848\pm 0.021$\\
\hline
Three loops & $0.0464\pm 0.0012$ & $20.860\pm 0.023$\\
\hline
Four loops & $0.0465\pm 0.0012$ & $20.861\pm 0.023$\\
\hline
\end{tabular}

\bigskip

2. The nonphysical cut is taken into account
, $\Lambda^{new}_3 = (1666\pm 7) MeV$. In this case the lower limit of the
integral in eq.(49) is equal to $-\Lambda^2_3$ and in the three-loop
approximaion $\Lambda_4=1591 MeV$, $\Lambda_5=791 MeV$, $\Lambda_6=280 MeV$.
Errors in this case are very small. The results of the calculations in the
three-loop approximation are the following:

$$
\alpha_s(-m^2_{\tau}) = 1.575, ~~ \alpha_s(m^2_{\tau} + i 0)=0.142+0.272 i,
~~ \alpha_s(-m^2_{\psi})=0.484 $$
$$ \alpha_s(m^2_{\psi}-0+i 0)=0.179+0.209
i ~~ \alpha_s(m^2_{\psi}+0+ 0 i)=0.223+229 i, ~~
\alpha_s(-m^2_{\Upsilon})=0.253
$$
\be
\alpha_s(m^2_{\Upsilon}-0+i
0)=0.202+0.110 i, ~~ \alpha_s(m^2_{\Upsilon}+0+0 i)=0.189+0.093 i, ~~
\alpha_s(-m^2_z)=0.139
\ee
$$ \alpha_s(-m^2_z+i 0)=0.129+0.036 i, ~~
r(m^2_z)=0.0457, ~~ R_l=20.845 $$

3. The nonphysical cut is taken into account, $\Lambda^{conv.}_3 = (618\pm
29) MeV$. The results of the calculations in 1-4 loop approximaion are given
in Table 4.


\begin{center}
Table 4.
\end{center}

\vspace{3mm}

The calculation of $\alpha_s(q^2)$ at different $q^2$ in 1-4 loop
approximation. The matching of $r(q^2)$ at the masses of $J/\psi$,
$\Upsilon$ mesons and of $2m_{\tau}$ is performed. The contribution of
nonphysical cut is taken into acccount.

\vspace{3mm}

\noindent
\begin{tabular}{|l|l|l|l|l|l|l|l|}
\hline Approx. & $\Lambda_3/MeV$ & $\Lambda_4/MeV$ &
$\Lambda_5/MeV$ & $\Lambda_6/MeV$ & $\alpha_s(-m^2_{\tau})$ &
$\alpha_s(m^2_{\tau}+i 0)$ \\ \hline
One loop & $370\pm 19$ &
$296\pm 16$ & $211\pm 13$ & $102\pm 7$ & $0.445\pm 0.015$ &
0.$222+(0.222\pm 007) i$ \\ \hline
Two loops & $539\pm 25$ &
$416\pm 21$ & $275\pm 15$ & $111\pm 7$ & $0.371\pm 0.012$ &
$0.19+(0.174\pm 0.006) i$\\
Three loops & $618\pm 29$ & $475\pm
24$ & $301\pm 16$ & $96\pm 5$ & $0.354\pm 0.010$ &
$0.186\pm0.001+(0.162\pm 0.005) i$\\
Four loops & $720\pm 33$ &
$557\pm 28$ & $359\pm 20$ & $61\pm4$ & $0.375\pm 0.012$ &
$0.184\pm 0.001+(0.161\pm0.005) i$\\ \hline
\end{tabular}

\vspace{3mm}

\noindent
\begin{tabular}{|l|l|l|l|}
\hline Approx. & $\alpha_s(-m^2_{\psi})$ & $\alpha_s(m^2_{\psi}-0+i
0)$ & $\alpha_s(m^2_{\psi}+0+i 0)$\\ \hline
One loop & $0.33\pm
0.01$ & $0.212\pm0.002+(0157\pm 0.005) i$ & $0.222\pm
0.002+(0.148\pm 0.005) i$\\ \hline
Two loops & $0.277\pm 0.006$ &
$0.182\pm 0.002+(0.125\pm 0.004) i$ & $0.192\pm 0.002+(0.118\pm
0.004) i$\\ \hline
Three loops & $0.266\pm 0.006$ & $0.177\pm
0.001+(0.116\pm 0.004) i$ & $0.187\pm 0.002+(0.112\pm 0.003) i$\\
\hline
Four loops & $0.273\pm 0.006$ & $0.177\pm
0.001+(0.116\pm0.003) i$ & $0.187\pm 0.002\pm(0.111\pm 0.003) i$\\
\hline
\end{tabular}

\vspace{3mm}
\noindent
\begin{tabular}{|l|l|l|l|}
\hline Approx. & $\alpha_s(-m^2_{\Upsilon})$ &
$\alpha_s(m^2_{\Upsilon}-0 +i 0)$ & $\alpha_s(m^2_{\Upsilon}+0+i
0)$\\ \hline
One loop & $0.180\pm 0.003$ & $0.180\pm
0.002+(0.082\pm 0.002) i$ & $0.184\pm 0.002+(0.076\pm 0.002) i$\\
\hline
Two loops & $0.189\pm 0.003$ & $0.157\pm 0.002+(0.066\pm
0.002) i$ & $0.161\pm 0.002+(0.061\pm 0.002) i$\\ \hline
Three
loops & $0.184\pm 0.003$ & $0.154\pm 0.002+(0.063\pm 0.002) i$ &
$0.159\pm 0.002+(0.059\pm 0.001) i$\\ \hline Four loops &
$0.185\pm 0.003$ & $0.154\pm 0.002\pm(0.063\pm 0.002) i$ &
$0.159\pm 0.002+(0.059\pm 0.002) i$ \\ \hline
\end{tabular}

\vspace{3mm}
\noindent
\begin{tabular}{|l|l|l|l|l|}
\hline Approx. & $\alpha_s(-m^2_z)$ & $\alpha_s(m^2_z+i 0)$ &
$r(m^2_z)$ & $R_l$\\ \hline
One loop & $0.135\pm 0.001$ &
$0.186\pm0.002+(0.032\pm0.001) i$ & $0.0420\pm 0.0004$ &
$20.772\pm 0.008$\\ \hline
Two loops & $0.120\pm 0.001$ &
$0.113\pm 0.001+(0.0274\pm 0.0005) i$ & $0.0395\pm 0.0003$ &
$20.721\pm 0.007$\\ \hline
Three loops & $0.118\pm 0.001$ &
$0.112\pm0.001+(0.0267\pm0.0004) i$ & $0.0389\pm 0.0003$ &
$20.710\pm 0.007$\\ \hline
Four loops & $0.118\pm 0.001$ &
$0.111\pm 0.001+(0.0266\pm 0.004) i$ & $0.0388\pm 0.0003$ & $20.708\pm
0.007$\\ \hline
\end{tabular}

\vspace{5mm}

Unlike conventional matching procedure at negative $q^2$, the
matching of $r(q^2)$ on the masses of $J/\Psi$, $\Upsilon$ mesons and of
$2 m_t$ is performed. The value $\alpha_s(-m^2_z)$ is practically
independent of the matching procedure. I believe, that the first alternative
is the best.


\section{ Comparison of the calculated values $R_T(s)$ with the
measured values $R_E(s)$.}


The value $R_T(s)$ is calculated by the formula

\be
R_T(s) = 3~ \sum~ e^2_q(1 + r(s))
\ee
The results of the calculations of $R(s)$ in three loop-approximation and of
their comparison with experiments are given in Table 5 for $2 \leq \sqrt{s}
\ll 4.8 GeV$ and for $12 \leq \sqrt{s} \leq 46.6~ GeV$ in Table 6. The
calculated values of the function $R(s)$ are in an excellent agreement with
the experiment except for the resonance region $3.7 \leq \sqrt{s} \leq 4.4~
GeV$. But the accuracy of measurements of $R(s)$ is insufficient to define
the value $r(s)$ with a good accuracy.


\bigskip

\centerline{\bf Table 5}
\begin{center}
Comparison of the calculated values $R_T(s)$ with the measured values $R_E$
[21]

\vspace{5mm}
\begin{tabular}{|c|c|c|c|c|c|}
\hline
$E_{cm}$ (GeV) & $R_T$ & $R_E$ & $E_{cm}$(GeV) & $R_T$ & $R_E$ \\
\hline
2.000 & 2.29 & $2.18.\pm 0,07\pm 0.18$ & 4.033 & 3.70 & $4.32\pm 0.17 \pm
0.22$\\
\hline
2.200 & 2.28 & $2.38 \pm 0.07\pm 0.17$ & 4.040 & 3.70 & $4.40 \pm 0.17 \pm
0.19$\\
\hline
2.400 & 2.27 & $2.38\pm 0.07\pm 0.14$ & 4.050 & 3.70 & $4.23\pm 0.17\pm
0.22$\\
\hline
2.500 & 2.27 & $2.39\pm 0.08\pm 0.15$ & 4.060 & 3.70 & $4.65\pm 0.19\pm
0.19$\\
\hline
2.600 & 2.26 & $2.38\pm 0.06\pm 0.15$ & 4.070 & 3.70 & $4.14\pm 0.20\pm
0.19$\\
\hline
2.700 & 2.26 & $2.30\pm 0.07\pm 0.13$ & 4.080 & 3.70 & $4.24\pm 0.21\pm
0.18$\\
\hline
2.800 & 2.25 & $2.27\pm 0.06\pm 0.14$ & 4.090 & 3.69 & $4.06\pm
0.17\pm 0.18$\\
\hline
2.900 & 2.25 & $2.22\pm 0.07\pm 0.13$ & 4.100 & 3.69 &
$3.97\pm 0.16\pm 0.18$\\
\hline
3.000 & 2.25 & $2.21\pm 0.05\pm 0.11$ & 4.110
& 3.69 & $3.92\pm 0.16\pm 0.19$\\
\hline
3.700 & 3.71 & $2.23\pm 0.08\pm 0.08$
& 4.120 & 3.69 & $4.11\pm 0.24\pm 0.23$\\
\hline
3.730 & 3.71 & $2.10\pm
0.08\pm 0.14$ & 4.130 & 3.69 & $3.99\pm 0.15\pm 0.17$\\
\hline
3.750 & 3.71
& $2.47\pm 0.09\pm 0.12$ & 4.140 & 3.69 & $3.83\pm 0.15\pm 0.18$\\
\hline
3.760 & 3.71 & $2.77\pm 0.11\pm 0.13$ & 4.150 & 3.69 & $4.21\pm 0.18\pm
0.19$\\
\hline
3.764 & 3.71 & $3.29\pm 0.27\pm 0.29$ & 4.160 & 3.69 & $4.12\pm
0.15\pm 0.16$\\
\hline
3.768 & 3.71 & $3.80\pm 0.33\pm 0.25$ & 4.170 & 3.69 & $4.12\pm 0.15\pm
0.19$\\
\hline
3.770 & 3.71 & $3.55\pm 0.14\pm 0.19$ & 4.180 & 3.69 & $4.18\pm 0.17\pm
0.18$\\
\hline
3.772 & 3.71 & $3.12\pm 0.24\pm 0.23$ & 4.190 & 3.69 & $4.01\pm 0.14\pm
0.14$\\
\hline
3.776 & 3.71 & $3.26\pm 0.26\pm 0.19$ & 4.200 & 3.69 & $3.87\pm 0.16\pm
0.16$\\
\hline
3.780 & 3.71 & $3.28\pm 0.12\pm 0.12$ & 4.210 & 3.69 & $3.20\pm 0.16\pm
0.17$\\
\hline
3.790 & 3.71 & $2.62\pm 0.11\pm 0.10$ & 4.220 & 3.69 & $3.62\pm 0.15\pm
0.20$\\
\hline
3.810 & 3.71 & $2.38\pm 0.10\pm 0.12$ & 4.230 & 3.69 & $3.21\pm 0.13\pm
0.15$\\
\hline
3.850 & 3.70 & $2.47\pm 0.11\pm 0.13$ & 4.240 & 3.69 & $3.24\pm 0.12\pm
0.15$\\
\hline
3.890 & 3.70 & $2.64\pm 0.11\pm 0.15$ & 4.245 & 3.69 & $2.97\pm 0.11\pm
0.14$\\
\hline
3.930 & 3.70 & $3.18\pm 0.14\pm 0.17$ & 4.250 & 3.69 & $2.71\pm 0.12\pm
0.13$\\
\hline
3.940 & 3.70 & $2.94\pm 0.13\pm 0.19$ & 4.255 & 3.69 & $2.88\pm 0.11\pm
0.14$\\
\hline
3.950 & 3.70 & $2.97\pm 0.13\pm 0.17$ & 4.260 & 3.69 & $2.97\pm 0.11\pm
0.14$\\
\hline
3.960 & 3.70 & $2.79\pm 0.12\pm 0.17$ & 4.265 & 3.69 & $3.04\pm 0.13\pm
0.14$\\
\hline
3.970 & 3.70 & $3.29\pm 0.13\pm 0.13$ & 4.270 & 3.69 & $3.26\pm 0.12\pm
0.17$\\
\hline
3.980 & 3.70 & $3.13\pm 0.14\pm 0.16$ & 4.280 & 3.69 & $3.08 \pm 0.12\pm
0.15$\\
\hline
3.990 & 3.70 & $3.06\pm 0.15\pm 0.18$ & 4.300 & 3.69 & $3.11\pm 0.12\pm
0.12$\\
\hline
4.000 & 3.70 & $3.16\pm 0.14\pm 0.15$ & 4.320 & 3.69 & $2.96\pm 0.12\pm
0.14$\\
\hline
4.010 & 3.70 & $3.53\pm 0.16\pm 0.20$ & 4.340 & 3.69 & $3.27\pm 0.15\pm
0.18$\\
\hline
4.020 & 3.70 & $4.43\pm 0.16\pm 0.21$ & 4.350 & 3.69 & $3.49\pm
0.14\pm 0.14$\\
\hline
4.027 & 3.70 & $4.58\pm 0.18\pm 0.21$ & 4.360 & 3.68
& $3.47\pm 0.13\pm 0.18$
\\
\hline
4.030 & 3.70 & $4.58\pm 0.20\pm 0.23$ &
4.380 & 3.68 & $3.50\pm 0.15\pm 0.17$\\
\hline
\end{tabular}
\end{center}

\newpage
\begin{center}
{\bf Table 5.~~ Continuation}
\end{center}

\begin{center}
\begin{tabular}{|r|r|r|}
\hline
$E_{cm}$ (GeV) & $R_T$ & $R_E$\\
\hline
4.390 & 3.68 & $3.48\pm 0.16 \pm 0.16$\\
\hline
4.400 & 3.68 & $3.91 \pm 0.16 \pm 0.19$\\
\hline
4.410 & 3.68 & $3.79 \pm 0.15 \pm 0.20$\\
\hline
4.420 & 3.68 & $3.68 \pm 0.14 \pm 0.17$\\
\hline
4.430 & 3.68 & $4.02 \pm 0.16 \pm 0.20$\\
\hline
4.440 & 3.68 & $3.85 \pm 0.17 \pm 0.17$\\
\hline
4.450 & 3.68 & $3.75 \pm 0.15 \pm 0.17$\\
\hline
4.460 & 3.68 & $3.66 \pm 0.17 \pm 0.16$\\
\hline
4.480 & 3.68 & $3.54 \pm 0.17 \pm 0.18$\\
\hline
4.500 & 3.68 & $3.49 \pm 0.14 \pm 0.15$\\
\hline
4.520 & 3.68 & $3.25 \pm 0.13 \pm 0.15$\\
\hline
4.540 & 3.68 & $3.23 \pm 0.14 \pm 0.18$\\
\hline
4.560 & 3.68 & $3.62 \pm 0.13 \pm 0.16$\\
\hline
4.60 & 3.68 & $3.31 \pm 0.11 \pm 0.16$\\
\hline
4.80 & 3.67 & $3.66 \pm 0.14 \pm 0.19$\\
\hline
\end{tabular}
\end{center}

\bigskip
\centerline{\bf Table 6.}


\begin{center}
Comparison of the calculated values $R_T(s)$ with the measured values $R_E$
[22-24]

\vspace{5mm}

\begin{tabular}{|c|c|c|c|c|c|}
\hline
$s/Gev$ & $R_T$ & $R_E$ & $\sqrt{s}/GeV$ &  &  \\
\hline
              &       &  [22]      &       &      &                     \\
\hline
14.0 & 3.92 & $4.10\pm 0.11\pm 0.11$& 29.93 & 3.88 &$3.55\pm 0.40\pm 0.11$\\
\hline
22.0 & 3.89 & $3.86\pm 0.12\pm 0.11$ & 30.38 & 3.87 & $3.85\pm 0.19\pm
0.12$\\
\hline
33.8 & 3.87 & $3.74\pm 0.10\pm 0.10$ & 31.29 & 3.87 & $3.83\pm 0.28\pm
0.11$\\
\hline
38.3 & 3.86 & $3.89\pm 0.10\pm 0.09$ & 33.89 & 3.87 &
$4.16\pm 0.10\pm 0.12$\\
\hline
41.5 & 3.86 & $4.03\pm 0.17\pm 0.10$ & 34.50 & 3.87 & $3.93\pm 0.20\pm
0.12$\\
\hline
43.5 & 3.86 & $3.97\pm 0.08\pm 0.09$ & 35.01 & 3.87 & $3.93\pm 0.10\pm
0.12$\\
\hline
44.2 & 3.86 & $4.01\pm 0.10\pm 0.08$ & 34.45 & 3.87 & $3.93\pm 0.18\pm
0.12$\\
\hline
46.0 & 3.86 & $4.09\pm 0.21\pm 0.10$ & 36.38 & 3.87 & $3.71\pm 0.21\pm
0.11$\\
\hline
46.6 & 3.86 & $4.20\pm 0.36\pm 0.10$ & 40.32 & 3.86 & $4.05\pm 0.19\pm
0.14$\\
\hline
     &       &  [23]                  &  41.18 & 3.86 & $4.21\pm 0.22\pm
     0.14$\\
\hline
29   & 3.88 & $3.96\pm 0.09$ & 42.55 & 3.86 & $4.20\pm 0.22\pm 0.14$\\
\hline
     &      &    [24]        & 43.53 & 3.86 & $4.00\pm 0.20\pm 0.14$\\
\hline
12   & 3.93 & $3.45\pm 0.27\pm 0.13$ & 44.41 & 3.86 & $3.98\pm 0.20\pm
0.14$\\
\hline
14.04 & 3.92 & $3.94\pm 0.14\pm 0.14$ & 45.59 & 3.86 & $4.40\pm 0.22\pm
0.15$\\
\hline
22    & 3.89 & $4.11\pm 0.13\pm 0.12$
 & 46.47 & 3.86 & $4.04\pm 0.24\pm
0.14$\\
\hline
25.01 & 3.88 & $4.24\pm 0.29\pm 0.13$ &        &      &             \\
27.66 & 3.88 & $3.85\pm 0.48\pm 0.12$ &       &      &              \\
\hline
\end{tabular}

\end{center}


\section{COMPARISON WITH $\alpha_s(s)$ OBTAINED FROM THE SUM RULES}


Perturbative corrections  to two measurements, namely, the Gross-Llewellyn
Smith sum rule [7] for the deep inelastic neutrino scattering and the
Bjorken sum rule [8] for polarized structure functions, have been
determined:

\be
\alpha_s(- 3 GeV^2) = 0.28 \pm 0.035 (stat) \pm 0.050 (sys.)~ [12,25]
\ee
and
\be
\alpha_s(-2.5 GeV^2) = 0.375^{+0.062}_{-0.081}~[12,26-28]
\ee
The results of the calculations give
\be
\alpha_s (- 3 GeV^2) = 0.381\pm 0.013
\ee

\be
\alpha_s( - 2.5 GeV^2) = 0.39\pm 0.013
\ee


\section{ Calculation of $R_l$}

\vspace{5mm}
The value $R_l = \Gamma(Z \to hadrons)/ \Gamma(Z \to leptons)$ is
parametrized by the latest  version of ZFITTER [29]

\be
R_l = 19.934 \Biggl ( 1 + 1.045 (\frac{\alpha_s}{\pi}) + 0.94
(\frac{\alpha_s}{\pi})^2 - 15 (\frac{\alpha_s}{\pi})^3 \Biggr )
\ee
This is the conventional method. In this parametrization

$$
M_H = 300 GeV, ~~ M_t = 174.1 GeV, ~~ 0.1 < \alpha_s < 0.13.
$$

In the method under investigation Eq.(65 ) should be changed by the formula

\be
R_l = 19.934 (1 + r(m^2_Z))
\ee
The calculated values of $r(m^2_z)$ and of $R_l$ are presented in Tables 3,4
in 1-4 loop approximation. The value $R_l$ can be compared with the
measurement [12].

\be
\Gamma(Z \to hadrons) = (1744.4 \pm 2)~MeV
\ee

\be
\Gamma(Z \to leptons) = (83.984 \pm 0.086)~ MeV
\ee

and

\be
R_l = 20.771 \pm 0.045
\ee
The value $R_l$ (69) does not contradict $R_l$ from Tables 3,4.


\begin{center}
\section{ On analiticity of $\alpha_s(q^2)$}
\end{center}


In the interesting papers [30] the renormgroup was combined with
analiticity of $\alpha_s(q^2)$.  It was assumed that $\alpha_s(q^2)$ is an
analytic function of $q^2$ in the whole complex $q^2$ plane with a cut along
the poitive $q^2$ semiaxis. In particular, it was obtained that
$\alpha_s(0)$ is universal and independent of the number of loops,
$\Lambda_3$ and

\be
\alpha_s(0) = 4 \pi/\beta^{(3)}_0 = 1.396
\ee
The analogous formula had been also obtained in paper [1]. As is seen from
Table 3, eq.(70) is valid only in one-loop approximation.

If $\alpha_s(q^2)$ has correct  analiticity, the Adler function will have
correct analiticity too, but a reverse statement is invalid.  If Adler
function has correct analiticity, function $a(q^2)$ may have, generally
speaking, additional singularities. This statement will be evident, if one
considers eq.(13) as an equation relative to $a(q^2)$. If only eq.(13) is
solved by expansion in $a(q^2)$, analiticity $a(q^2)$ will be the same as the
Adler function. However, using of $d(q^2)$ expansion in $a(q^2)$ at small
$q^2$ seems to be doubtful.


\section{Conclusion.}


Concluding, let us formulate the main results of this paper.

1) There are two and only two values of $\Lambda_3$, at which $R_{\tau,V+A}
= 3.475\pm 0.022$, one conventional value $\Lambda^{conv}_3
= (618\pm 29)~MeV$ ($\Lambda_3$ is defined by eq.(7)) and the other, found in
this paper, $\Lambda^{new}_3 = (1666\pm 7)~MeV$.

2. The renormgroup calculation leads to appearance of a nonphysical cut in
the Adler function. In complete theory, where everything is taken into
account, the nonphysical cut must be absent. The question arises, if it is
possible to neglect nonphysical cut at the present situation with the theory.
At conventional value $\Lambda_3=\Lambda^{conv}_3$ it is impossible. The new
value, $\Lambda_3=\Lambda^{new}_3$ is more preferable than
$\Lambda_3=\Lambda^{conv}_3$, since at $\Lambda_3=\Lambda^{new}_3$ the
contribution of nonphysical cut into $R_{\tau,V+A}$ is practically absent.

3) At $\Lambda_3=\Lambda^{conv}_3$ there is an essential disagreement
between the ALEPH experiment and new obtained sum rules, which follow only
from analytical properties of the polarization operator. This disagreement
disappears if $\Lambda_3=\Lambda^{new}_3$.

4) In 1-4 loop approximation all calulations are made exactly, without
$\pi/ln \frac{Q^2}{\Lambda^2}$ expansion.

5) At $\Lambda^{new}_3 = (1565\pm 193)~ MeV$ the nonphysical cut may be
omitted, the polarization operator has correct analytical properties and
$R_{\tau,V+A}=3.475\pm 0.022$. But in this case QCD parameters  essentially
differ from conventional values.


\centerline{\bf Acknowledgements}

In conclusion, thanks are due to B.L.Ioffe and K.N.Zyablyuk for useful
discussions and to K.N.Zyablyuk for checking the correctness of tables 1,2.
The author is also indebted to M.Davier for his kind presentation of the
ALEPH experimental data.

\end{document}